\documentclass[fleqn,usenatbib]{mnras}

\usepackage{newtxtext,newtxmath}
\usepackage[]{algorithm2e}

\usepackage[T1]{fontenc}
\usepackage{ae,aecompl}
\usepackage{pifont}

\usepackage{graphicx}	
\usepackage{amsmath}	
\usepackage{amssymb}	

\newcommand{\kms} {$\mathrm{ km \; s^{-1}}\,$}
\newcommand{\msol} {M$_{\odot}$}
\newcommand{\lsol} {L$_{\odot}$}
\newcommand{\tsol} {T$_{\odot}$}

\newcommand{\zsol} {Z$_{\odot}$}

\newcommand{\about} {$\sim$}

\newcommand{\halpha} {$\mathrm{H\alpha}$ }
\newcommand{\hbeta} {$\mathrm{H\beta}$ }
\newcommand{\ang} {\r{A} }

\newcommand{\hoki}{{\tt hoki}\,}

\title[A Systematic ageing Method I]{A systematic ageing method I:\\ H\,II regions D118 and D119 in NGC 300}

\author[H.F. Stevance]{H. F. Stevance$^{1}$\thanks{E-mail: hfstevance@gmail.com}, J.J. Eldridge$^{1}$, A. McLeod$^{2}$, E. R. Stanway$^{3}$ , A. A. Chrimes$^{3, 4}$  \\
$^{1}$University
 of Auckland, Department of Physics and Astronomy, 38 Princes Street, 1142, Auckland, New Zealand.\\
$^{2}$Department of Astronomy, University of California Berkeley, Berkeley, CA 94720, USA\\
$^{3}$Department of Physics, University of Warwick, Gibbet Hill Road, Coventry, UK\\
$^{4}$Department of Astrophysics/IMAPP, Radboud University, Nijmegen, The Netherlands\\
}

\date{Accepted XXX. Received YYY; in original form ZZZ}

\pubyear{2020}

\begin{document}
\label{firstpage}
\pagerange{\pageref{firstpage}--\pageref{lastpage}}
\maketitle

\begin{abstract}
Accurately determining the age of H\,{\sc ii} regions and the stars they host is as important as it is challenging. 
Historically the most popular method has been isochrone fitting to Hertzsprung-Russell Diagrams or Colour-Magnitude Diagrams.
Here we introduce a different method for age determination using BPASS and \hoki. 
We infer the most likely ages of the regions D118 and D119 NGC 300 to be log(age/years)=$6.86^{+0.05}_{-0.06}$  and we also deduce stellar mass and number counts by comparison with the BPASS models. 
We compare how our binary and single star models perform and find that the latter are unable to predict 20 per cent ($\pm10$ per cent) of our sample. 
We also discuss how results obtained from isochrone fitting would differ. 
We conclude that ages could be underestimated by \about0.2 dex and that the limitations of the isochrone method is not solely due to the lack of binary stars. 
We propose that the method presented here is more reliable and more widely applicable since it can be used on smaller samples. 
Alongside this study, we release new \hoki features to allow easy implementation of this method.
\end{abstract} 



\defcitealias{mcleod20}{M20}. 


\section{Introduction}

The study of H\,{\sc ii} regions is a crucial probe of recent star formation and young stellar populations, but accurately ageing H\,{\sc ii} regions and their components is as important as it is challenging.
One of the most common methods consists in visual comparison of Hertzsprung-Russell Diagrams (HRDs) or Colour-Magnitude Diagrams (CMDs) of observed clusters to Main-Sequence or pre-Main Sequence isochrones obtained from theory (\citealt{vazquez96, walborn97, massey98, melena08, wright10, wang11}). 
If spectroscopic observations are available, the age of individual stars can also be deduced if by comparing their spectral class to evolutionary tracks (e.g \citealt{wright10}) and the [O \,{\sc iii}]/\hbeta line indicator can be used to globally age an H\,{\sc ii} region (e.g. \citealt{copetti86, dors08}).

The most commonly used evolutionary tracks and isochrones unfortunately do not take into account the effects of interacting binaries (e.g \citealt{schaller92, siess00, meynet03}). 
This is particularly problematic since H\,{\sc ii} regions host massive stars and we know that most of these will be in binary systems \citep{sana12, moe17}.
The inclusion of binaries allows a given stellar population to stay hotter for longer and to produce harder ionizing radiation, as a direct result of binary interaction 
\citep{2016MNRAS.456..485S,eldridge17}:
This affects the path that a stellar population will follow on the HRD or CMD and how long it will spend in a given region.
Therefore, this could significantly affect age determination performed by comparison to single star isochrones. 

Isochrone studies of individual H\,{\sc ii} regions have found that their age ranges from 1 to 10 Myrs, with many clusters younger than 4 Myrs (e.g. Trumpler 14 - \citealt{vazquez96}; Trumpler 15 - \citealt{wang11}; NGC 3603 - \citealt{melena08}; 30 Dor - \citealt{walborn97}; Cyg OB2 - \citealt{wright10}; Westerlund 1 - \citealt{aghakhanloo20}).
Isochrone fitting has two main drawbacks: (i) it cannot take into account the effects of binary interactions and (ii) The large data sets  it typically requires (from a couple dozens to as many hundreds of sources, e.g. \citealt{melena08, wright10}) may not always be available for distant or low mass clusters.
Alternatives to isochrone fitting like the Red Supergiant method \citep{britavsky19, beasor19}, also require that specific conditions be met (e.g. the presence of Red Supergiants) to be applied. 

In this work we introduce a method for age determination based on the comparison of the Binary Population and Spectral Synthesis (BPASS) models\footnote{Available at \url{https://bpass.auckland.ac.nz} and \url{https://warwick.ac.uk/bpass}.} to observed Hertzsprung-Russell Diagrams (HRDs) or Colour-Magnitude Diagrams (CMDs).
It can be systematically applied to resolved star populations with any size sample, at any age, and which takes into account binary systems which are particularly important in massive star populations \citep{sana12}.
New tools and the {\tt AgeWizard} pipeline are made available in {\tt hoki}\citep{stevance20} \footnote{{\label{github}}https://github.com/HeloiseS/hoki} to allow the community to easily apply this method to any sample. 

In this paper, we focus on the HRDs H\,{\sc ii} regions located in NGC 300 (\about2Mpc, \citealt{dalcaton09}).
Recently, \citet[][hereafter \citetalias{mcleod20}]{mcleod20} extracted individual stellar spectra in NGC 300 by combining Integral Field Unit (IFU) MUSE data and high resolution Hubble Space Telescope (HST) photometry. 
This novel technique unlocks access to stellar parameters of individual sources outside of the Milky Way and Magellanic Cloud systems, where blending of individual stars traditionally limited ground-based spectroscopic observations. 

In Section \ref{sec:obs_all} we provide a brief introduction to BPASS, we summarise the data set produced by \citetalias{mcleod20} for the H\,{\sc ii} region complexes D118 and D119 and we describe our age determination method. 
In Section \ref{sec:hrds} we determine the ages of individual cluster members and combine these estimates to infer the most probable age of D118 and D119.
In Section \ref{sec:sec4} we compare the observed number of stars and combined ionizing flux to that of BPASS. 
We deduce a cluster mass and a star count at the preferred ages of D118 and D119. 
Two independent age estimates are also performed.
In Section \ref{sec:discussion} we discuss our age and mass estimates and explore how isochrone fitting and the use of single star models can affect age determination. 
Finally we summarise our results in Section \ref{sec:conclusion}.

\section{Observations and models}\label{sec:obs_all}
\subsection{Observational Data}
\label{sec:obs}
The stars considered in this work (see Table \ref{tab:obs}) belong to two giant H\,{\sc ii} region complexes --  D118 and D119 -- located within NGC 300 (D\about2 Mpc, Z=0.33\zsol). 
We remark that the D118 and D119 notation is an abbreviation of the format established by \cite{deharveng88} which was introduced in \citetalias{mcleod20}.
For consistency with table 3 of \citetalias{mcleod20}, however,  we conserve the original nomenclature in our Table \ref{tab:obs}. 
There are five individual H\,{\sc ii} regions identified by \cite{deharveng88} and analysed by \citetalias{mcleod20}: 118-A, 118-B, 119-A, 119-B, 119-C.
For the majority of this study however, we will consider the D118 and D119 ensembles rather than their individual components. 
This is because the sample size is limited, and their proximity suggests that they are likely to be coeval within the limit of the time resolution (0.1 dex) of our models. 

The H\,{\sc ii} regions were observed by \citetalias{mcleod20} using the Multi Unit Spectroscopic Explorer (MUSE -- \citealt{bacon10}) on the Very Large Telescope, and high angular resolution Hubble Space Telescope catalogues were used to de-blend and resolve single stars. 
Further details on the observations and the data reduction can be found in Section 2 of \citetalias{mcleod20}.

In order to derive stellar parameters, \citetalias{mcleod20} fitted the extracted spectra of each star to PoWR \citep{hainich19} atmosphere model grids.
As they were solely calculated for Galactic, LMC and SMC metallicities, and given that for NGC 300 $Z\approx0.33$\zsol\ \citep{butler04}, the LMC PoWR grids ($Z=0.5$\zsol) were the most consistent.
The best fit were identified using $\chi^2$ minimisation. 
One major source of uncertainty is that unresolved binaries are not taken into account -- photon fluxes calculated therefore provide an upper limit on the stellar mass.
Additionally, the limited number of strong WR star features in the spectra of two of the 4 WR stars observed by \citetalias{mcleod20} led to unconstrained stellar parameters. 
Therefore they cannot be used in this work. 

In the case of 119-3, the stellar parameters derived from the initial PoWR model fits yielded a mass (15.7\msol -- see table 2 of \citetalias{mcleod20}) that is too low to explain the prominent He\,{\sc ii} $\lambda5411$ lines seen in the spectrum of this source. 
Focusing on the He\,{\sc ii} $\lambda6406$ line for their PoWR fits, they re-evaluate the luminosity and mass of 119-3.
In this work we consider both sets of stellar parameters: 119-3 and 119-3b, corresponding to the original and revised parameters, respectively. 

\begin{table*}
\centering
\caption{\label{tab:obs} Stellar parameters for the O-type and WR stars considered in this work as obtained by \citetalias{mcleod20} (see their tables 3 and 4) from best-fit PoWR atmosphere models \citep{hainich19}. The most likely age of each star according to the probability density functions inferred from the synthetic HRDs of BPASS are also given -- the non-gaussian nature of most of these distributions does not allow us to provide errors bars and we refer the reader to Figure \ref{fig:stars_age_pdf}. Note that the 119-3 and 119-3b stellar parameters refer to the same source but correspond to two different PoWR model fits -- for further details see text in Section \ref{sec:obs}} 

\begin{tabular}{l c c c c c r r r}

\hline
ID & log(L)  & log(T) & \multicolumn{3}{c}{Most likely age} & \multicolumn{3}{c}{P($6.7\le$log(age/years)$\le6.9$)}\\
 & (\lsol) & (\tsol) & \multicolumn{3}{c}{log(years)} & & &  \\
 & & & $Z$=0.006 & $Z$=0.008 & $Z$=0.010 & $Z$=0.006 & $Z$=0.008 & $Z$=0.010  \\
\hline
[DCL88]118-1 & 5.0 & 4.48 & 6.9 & 6.8 & 6.8 & 86 \% & 78 \%  &  58  \% \\ 

[DCL88]118-2 & 5.1 & 4.45 & 6.7 & 6.7 & 6.7 & 93 \% & 74 \%  & 67 \%  \\

[DCL88]118-3 & 4.9 & 4.46 & 6.8 & 6.9 & 6.8 & 78 \% & 66 \%  & 56 \%  \\

[DCL88]118-4 & 5.9 & 4.47 & 6.5 & 6.5 & 6.5 & 13 \% & 13 \%  & 10 \%  \\

[DCL88]118-WR2 & 5.3 & 4.9 & 6.9 & 6.8 & 6.8 & 85 \% & 72 \%  & 96 \% \\

[DCL88]119-1 & 5.0  & 4.48 & 6.9 & 6.8 & 6.8 & 86 \% & 78 \%  & 58  \% \\

[DCL88]119-2 & 5.4 & 4.53 & 6.7 & 6.7 & 6.7 & 95 \% & 74 \%  &  58 \% \\

[DCL88]119-3 &  4.3 & 4.52 & 6.8 & 6.8 & 6.7 & 50 \% & 47 \%  & 35 \% \\

[DCL88]119-3b & 5.7 & 4.52 & 6.7 & 6.5 & 6.5 & 67 \% & 28 \%  & 17 \% \\

[DCL88]119-4 & 4.5 & 4.52 & 6.9 & 6.9 & 6.9 & 45 \% & 55 \%  & 53 \%  \\

[DCL88]119-5 & 4.5 & 4.56 & 7.3 & 7.3 & 7.3 & 2 \% & 2 \%  & 3 \% \\

[DCL88]119-6 & 4.9 & 4.46 & 6.8 & 6.9 & 6.8 & 78 \% & 66 \%  &  56 \% \\

[DCL88]119-7 & 4.5 & 4.52 & 6.9 & 6.9 & 6.9 & 45 \%  & 55 \%  & 53 \% \\

[DCL88]119-8 & 4.3 & 4.52 & 6.8 & 6.8 & 6.7 & 50 \% & 47 \%  & 35 \% \\

[DCL88]119-9 & 4.5 & 4.52 & 6.9 & 6.9 & 6.9  & 45 \%  &  55 \%  &  53 \%\\

[DCL88]119-WR1 & 5.3 & 4.65 & 6.9 & 6.9 & 6.9 & 78 \% & 79 \%  &  74 \% \\
\hline

\end{tabular}
\end{table*}

\subsection{Using BPASS and {\tt hoki} to determine ages}

\subsubsection{BPASS and \hoki}
\label{sec:model}

The BPASS models simulate the evolution of a stellar population from a single burst of star formation. They have been calculated for a range of metallicities and initial mass functions (IMFs) and have the ability to include a realistic population of binary stars. 
An exhaustive description of the physical prescriptions included in the models can be found in \cite{eldridge17, stanway18}.
In this work we use the outputs of BPASS v2.2.1 with the fiducial IMF defined as a broken power law with $dN/dM =-1.35$ between 0.1 and 0.5 \msol and $dN/dM =-2.35$ between 0.5 and 300 \msol \citep{kroupa93}, and a metallicity $Z= 0.006$ corresponding to $\approx$ 1/3 \zsol, which is the closest BPASS metallicity to NGC 300. 
We also consider $Z= 0.008$ and $Z= 0.010$ (i.e. 0.4\zsol\ and 0.5 \zsol) since the physical parameters of the stars in \citetalias{mcleod20} were found based on $Z=0.5$\zsol\ PoWR models. 

{\tt hoki}   is specifically designed to interface with BPASS outputs and offers tools to easily load modelled data, perform technical pre-processing in the background, and automate some of the fundamental steps of data analysis \citep{stevance20}.

Alongside this paper we make available \hoki v1.5 -- ``The {\tt AgeWizard} release'' -- which includes all the necessary tools to infer likely ages from comparison of observational data with BPASS HRDs and CMDs.
Our method is described in the following section.

\subsubsection{\tt AgeWizard}
\label{sec:mymethod}

Our age determination technique is based on the comparison of observational data to HRDs or CMDs modelled by BPASS.
BPASS provides HRDs (CMDs) at 51 time bins separated by 0.1 dex with 6.0$\le$ log(age/years)$\le$ 11.0.
A BPASS HRD/CMD is composed of a grid where each cell $i$ is filled with the number of stars at time $t$ $N_{i,t}$.

\begin{figure}
	\includegraphics[width=7cm]{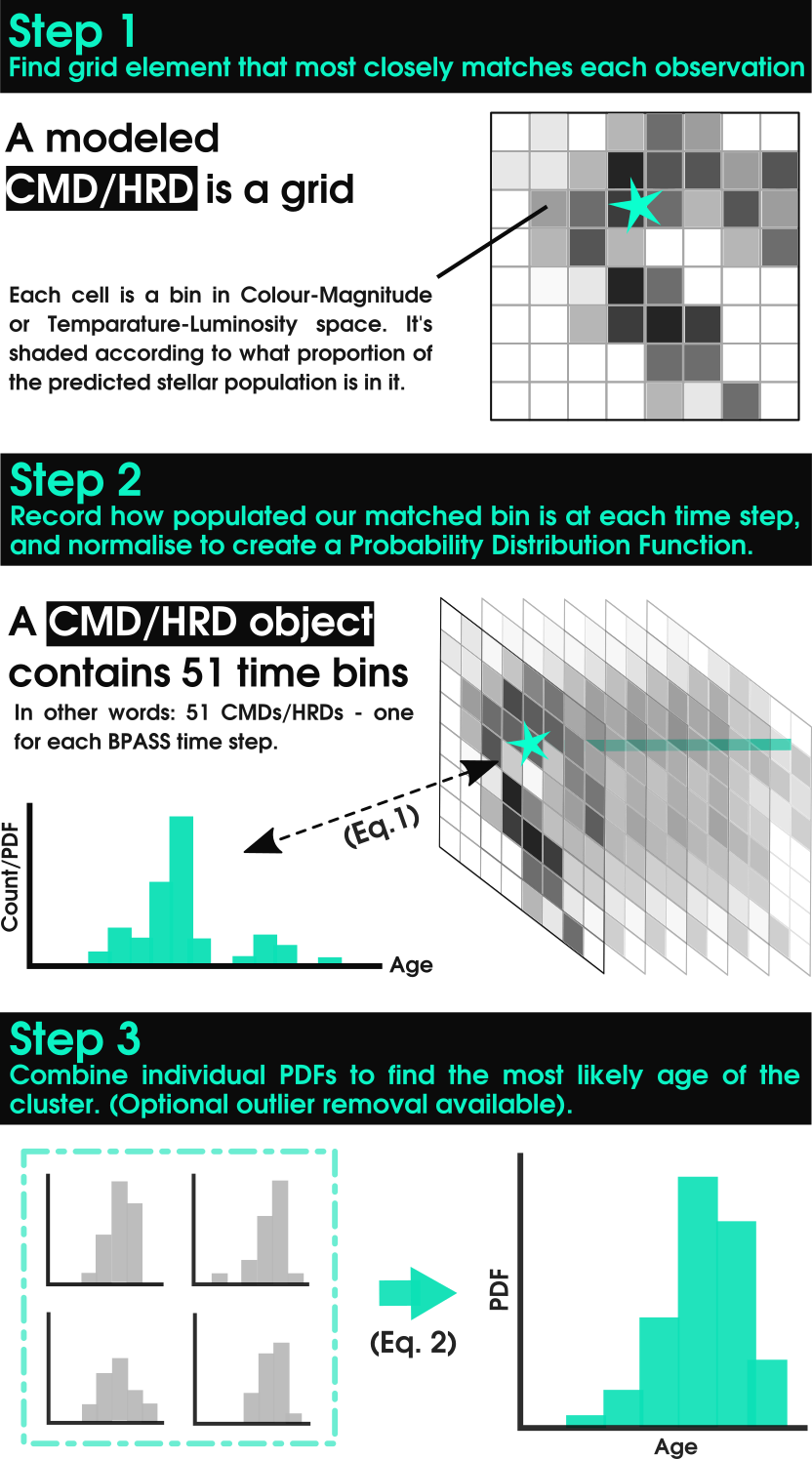}
    \caption{\label{fig:cartoon} Visual representation of the AgeWizard method.}
\end{figure}

For each source, the method is as follows (also see Figure \ref{fig:cartoon}):
\begin{itemize}
    \item \textbf{Step 1:} Identify the grid element $j$ that best matches the observations.
    \item \textbf{Step 2:} For each time step, record $N_{j,t}$ and calculate the age Probability Distribution Function (PDF) for each source/cell $j$ --$f_j(t)$.
    \item \textbf{Step 3:} Combine the PDFs of each cluster members to obtain the age PDF for the cluster --$f(t)$.
\end{itemize}

The age PDF of an individual cluster member $f_j(t)$ is the normalised distribution of counts recorded in cell $N_j$ at each time $t$:
\begin{equation}\label{eq:fi}
    f_j(t) = \frac{N_{j, t}}{\sum\limits_{t} N_{j, t}},
\end{equation}
and the age PDF of the whole population is the sum of the individuals age PDFs, once again normalised over time:
\begin{equation}\label{eq:f}
    f(t) = \frac{\sum\limits_{j}f_j(t)}{\sum\limits_{j, t} f_j(t)},
\end{equation}

In practice $f(t)$ may be better calculated by removing some sources $j$ from the sample if outliers are identified, such as foreground/background stars, etc... 
This is exemplified in the NGC 300 analysis presented below. It is also worth noting that if multiple observations fall within the same cell, that cell will be counted multiple times. Finally, we ensure that the age PDFs are 0 in age bins older than the current age of the universe.

Jupyter notebooks showing the complete analysis contained within this paper have been made available online for reproducibility purposes\footnote{https://github.com/UoA-Stars-And-Supernovae/A\_systematic\_aging\_method\_I}.

\section{Hertzsprung-Russell Diagrams}
\label{sec:hrds}
\subsection{Stellar age estimate from HRDs}
\label{sec:stellar_ages}

\begin{figure*}
	\includegraphics[width=17cm]{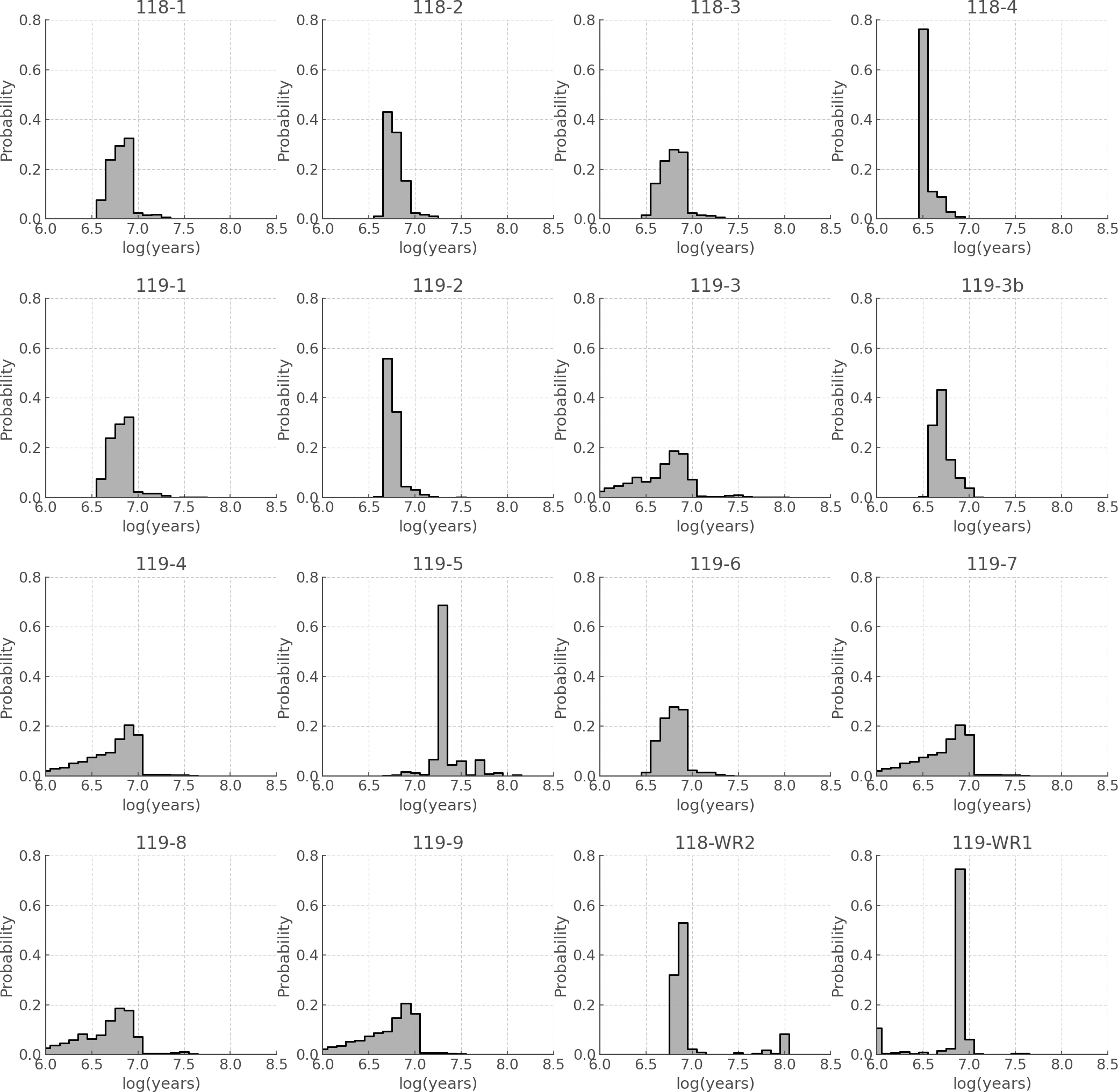}
    \caption{\label{fig:stars_age_pdf} probability density functions of the age of each star inferred from the stellar parameters quoted in Table \ref{tab:obs} and the synthetic HRDs of BPASS for a standard IMF and a metallicity $Z=0.006$. The most likely age of each star is summarised in Table \ref{tab:obs}. }
\end{figure*}

\begin{figure*}
	\includegraphics[width=17cm]{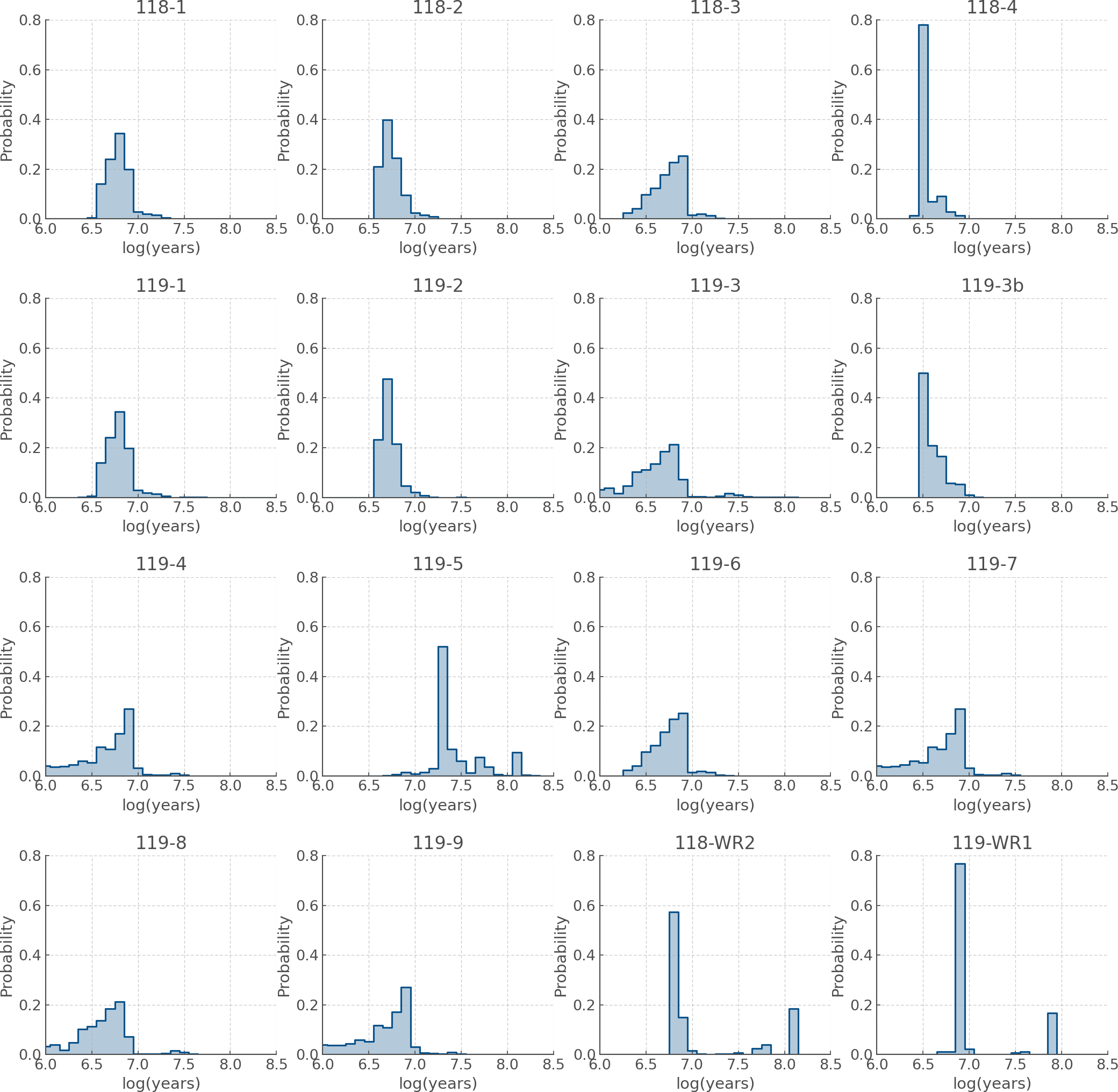}
    \caption{\label{fig:stars_age_pdf_z008} Same as Figure \ref{fig:stars_age_pdf} for a metallicity $Z=0.008$.}
\end{figure*}

\begin{figure*}
	\includegraphics[width=17cm]{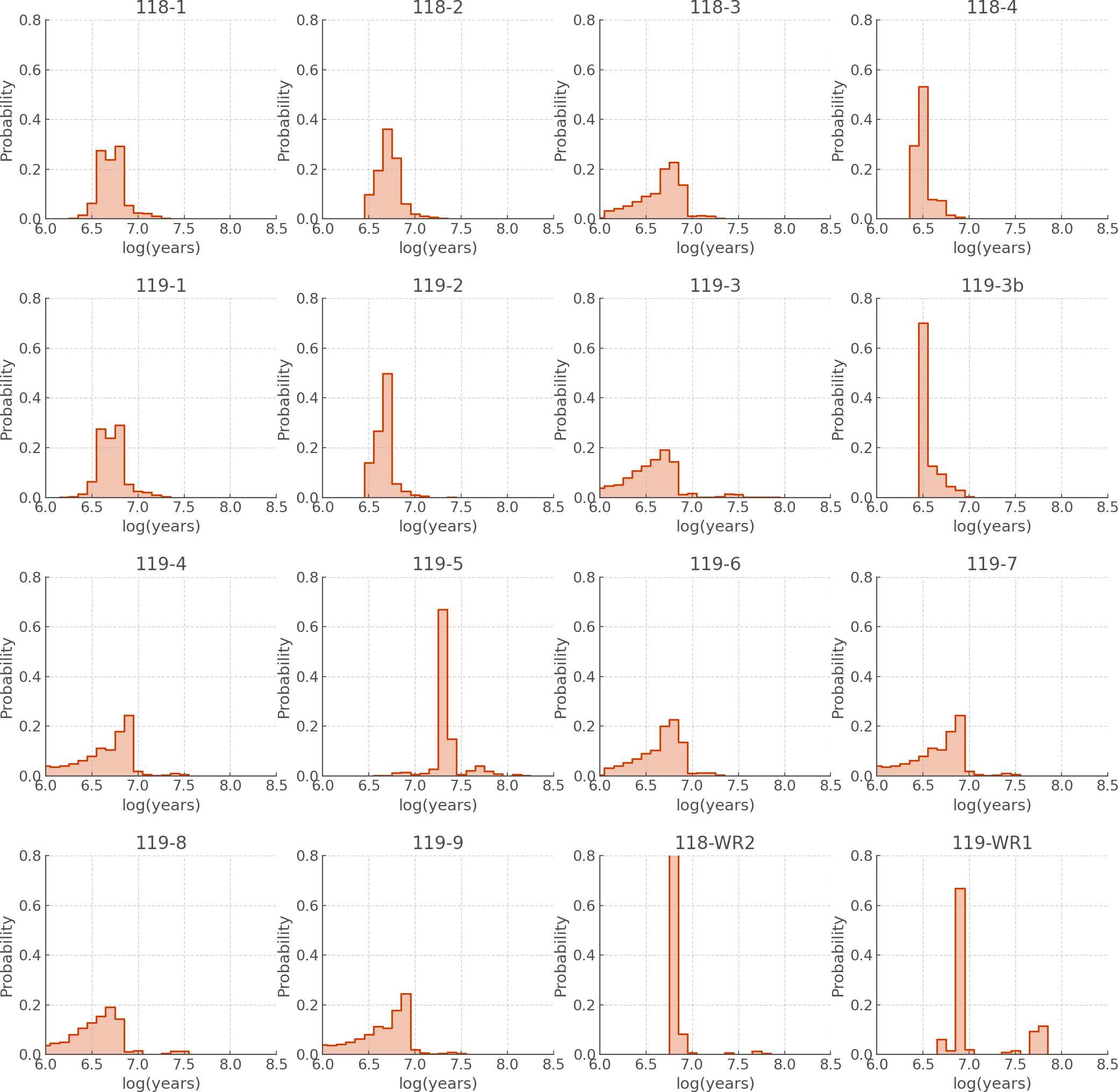}
    \caption{\label{fig:stars_age_pdf_z010} Same as Figure \ref{fig:stars_age_pdf} for a metallicity $Z=0.010$ or half solar (here we take solar metallicity to be $Z=0.020$).}
\end{figure*}

We apply the method described in \ref{sec:mymethod} to the sources listed in Table \ref{tab:obs}.
The PDFs $f_f(t)$ are shown in Figures \ref{fig:stars_age_pdf}, \ref{fig:stars_age_pdf_z008}  and \ref{fig:stars_age_pdf_z010}, for $Z=0.006, 0.008$ and 0.010 (respectively). 
The most likely age peak of the distribution) for each star is also summarised Table \ref{tab:obs}.

Overall, most stars have similar preferred ages: between 5 and 8 Myrs (log(age/years)=6.7-6.9).
At $Z=0.006$, only 118-4 and 119-5 deviate from this trend, being respectively younger (3 Myrs) and older (20 Myrs) than the rest of the population; the same most likely ages and very consistent PDFs are found for these sources at $Z=0.008$ and 0.010.
Additionally, at $Z=0.008$ and 0.010, 119-3b is also found to have a most likely age younger than the rest of the population (log(age/years)=6.5), whereas at $Z=0.006$ it is consistent with the majority of our sources. 

It is also interesting to note that the PDFs of at least one of the WR stars at all three metallicities shows a secondary probability peak between log(age/years)\about7.5 and log(age/years)\about8.0, which corresponds to 30 -- 100 Myrs. 
This is an unphysical age range for a WR star and it must therefore correspond to another type of source that is found in a similar region of the HRD as WR stars. 
In this case they are central stars of planetary nebulae, i.e. young white dwarfs towards the upper range of the maximum mass.

We will first discuss these outliers before attempting to combine the individual PDFs to infer the most likely age of D118 and D119.

\subsection{Outliers}
\label{sec:outliers}
\subsubsection{118-4}
As seen in the Figure  \ref{fig:stars_age_pdf}, the position of 118-4 on the HR diagram corresponds to a most likely age that is younger (log(age/years)=6.5, equivalent to \about 3 Myr) than all other stars in D118 and D119, which typically prefer ages around 5--8 Myrs. 
The asymmetric probability density functions of some stars in our sample do extend down to very early ages that overlap with 118-4 -- that is the case for 119-4,7,8 and 9.
However, \about 60-70 per cent of stars at the HRD location corresponding to these targets have a log(age/years) between 6.7 and 6.9 years, whereas at the HRD bin associated with 118-4, only 3.7 per cent of the sample falls in this age range.

Consequently, it is highly likely that 118-4 genuinely is, or \textit{appears}, younger.
If we consult \citetalias{mcleod20}, particularly their Figure 5, we can see that 118-4 is spatially separated from most stars in D118, which could indicate that it formed at a slightly later date. 
However, 118-2 and 118-WR2 are also somewhat isolated, and their most likely ages are consistent with the other stars in this region. 

Another way to explain the youthful appearance of 118-4 is through rejuvenation: if the star underwent a merger event, or accreted a substantial amount of material from a binary companion, it would present as a younger star. BPASS at the moment does have a simple inclusion of the effect of rejuvenation \citep{eldridge17}. 
The current BPASS implementation does not include the effects of rotation in extending stellar lifetimes at this metallicity: quasi-homogeneous evolution is only implemented up to $Z=0.004$ (which is clearly visualised in fig. 15 of \citealt{chrimes20}).
We note that \cite{demink13} found that any mass transfer in a binary can lead to the accretor being spun up to high rotation velocities, which (at low metallicities) could significantly extend the lifetime of massive stars.

The slight difference in the fitting of this star to the rest of the observed stellar population suggests that the rejuvenation model should be revisited.
This would involve including a more complex treatment of stellar rotation within the stellar models to extend the lifetime of the main-sequence star.
The fact that \citetalias{mcleod20} find this source to be the most luminous and most massive in their sample count hints that it could be the result of a stellar merger.

\subsubsection{119-3b}
The PDF of the age of the 119-3b stellar parameters evolve to prefer lower ages, $\log({\rm age/yr})=6.5$, as metallicity increases.
Overall, the PDF and preferred age of 119-3b are similar to 118-4.
This is consistent with the derived masses (41.3 and 49.2 \msol\ for 119-3b and 118-4, respectively -- see table 2 of \citetalias{mcleod20}), which are significantly higher than for the other sources in the sample. 
The greater mass of 118-4 compared to 119-3b also explains why the shape of the age probability density is greater towards lower ages. 
As in the case of 118-4, a merger and rejuvenation would explain the observed stellar parameters.

\subsubsection{119-5}
In contrast to 118-4, 119-5 appears much older than the other sources in our sample, with most likely log(age/years)=7.3  (\about 20 Myrs).
At such ages the gas will have dispersed and the nebular emission characteristic of H\,{\sc ii}  regions will have faded.
In Figure 9 of \citetalias{mcleod20} we can see that 119-5 is located away from the core of 119-A, which is inhabited by 119-3,\,4,\,6,\,8 and 9. 

A closer look at the BPASS models resulting in the PDF associated with this location on the HRD reveals that models for ages $<$10Myr all correspond to main sequence stars, whereas above 10~Myrs the sample is composed of stellar mergers from lower mass stars and secondary stars rejuvenated through binary interaction. 
In both cases, rapid rotation is expected to result from the interaction which would cause the spectral lines to appear broadened. 
That is not the case, however, as can be seen in Figure 4 of \citetalias{mcleod20}.
On the whole, although the probability distribution at this location of the HRD peaks at a later ages, it is more likely that 119-5 corresponds to one of the low probability age bins at log(Age/years)$\la$7.0.

\subsubsection{WR stars at 100 Myrs?}
As mentioned in Section \ref{sec:stellar_ages}, more than half of the age PDFs recovered for the WR star locations on the HRD show secondary probability peaks between log(Age/years)=7.5 and 8.0 (see Figures \ref{fig:stars_age_pdf}, \ref{fig:stars_age_pdf_z008} and \ref{fig:stars_age_pdf_z010}). 
This does not impact our ability to recover a sensible age for WR stars: at all metallicities both 118-WR2 and 119-WR1  have at least a \about 70 per cent (up to 97 per cent) probability of having an age between log(Age/years)=6.7 to 6.9 (\about 5--8 Myrs). 

The secondary probability surge at later ages can be explained by the presence of helium stars or the central stars of planetary nebulae in the BPASS models which are found in the same region of the HRDs but at later times. 
These therefore do not correspond to classical WR stars but will appear in the age PDFs corresponding to their stellar parameters log(T) and log(L).

\subsection{Aggregate age}
\label{sec:region_ages}
In order to estimate the age of D118, D119 and the entire region, we calculate $f_{D118}(t)$, $f_{D119}(t)$ and $f_{all}(t)$ according to Equation \ref{eq:f}.
We remark that 119-3 represents a different set of parameters for the same source as 119-3b, but as demonstrated in \citetalias{mcleod20}, the 119-3b stellar parameters are a more reliable choice for the target. 
Therefore we omit 119-3 from our samples before calculating $f_{D119}(t)$ 

\begin{figure*}
	\includegraphics[width=18cm]{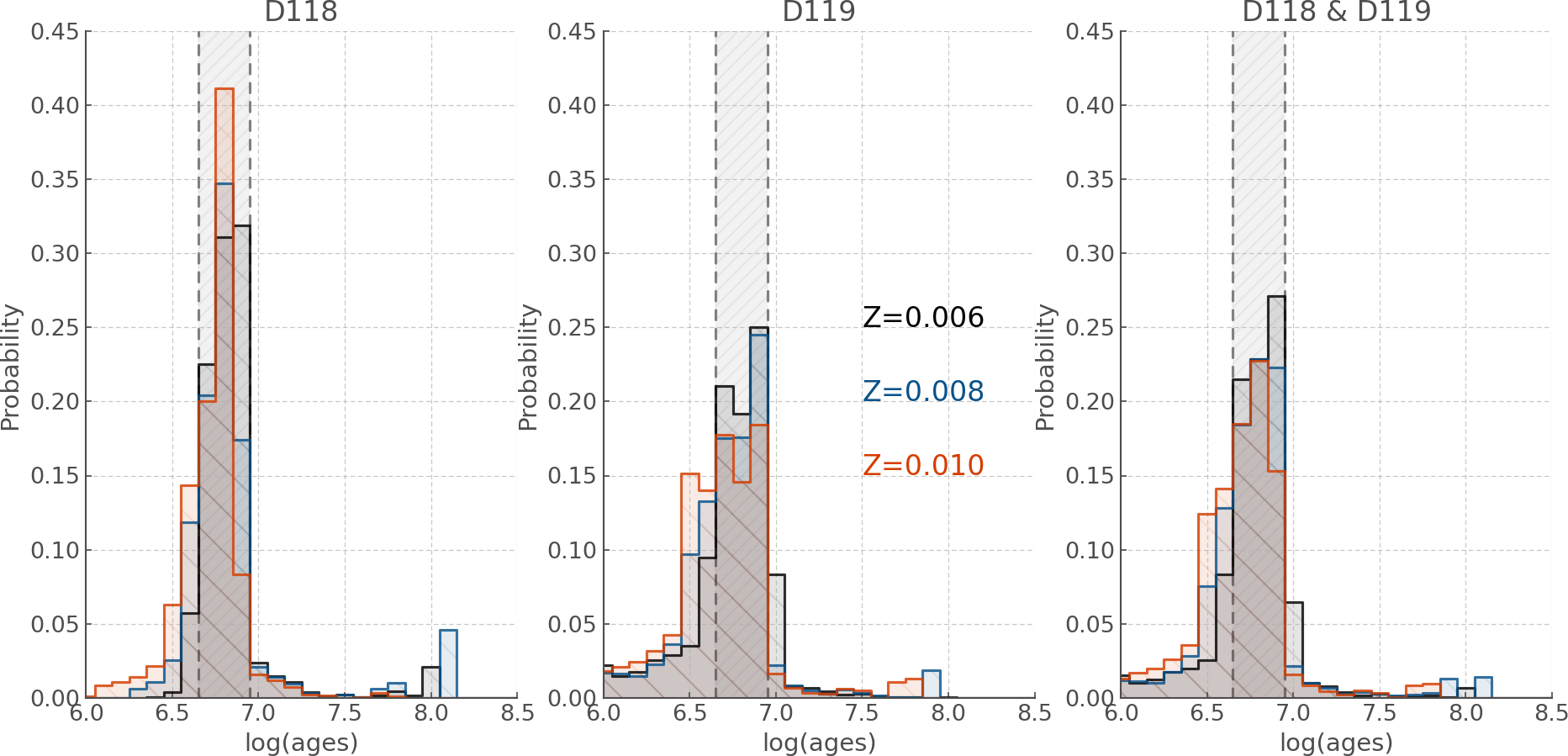}
    \caption{\label{fig:combined_age_pdf} Probability density functions of the age of D118 and D119 at Z=0.006 (black), Z=0.008 (blue), Z=0.010 (red). The favoured age range is highlighted by the dotted lines and grey shaded area. }
\end{figure*} 

In Figure \ref{fig:combined_age_pdf} we present the PDFs $f_{D118}(t)$, $f_{D119}(t)$ and $f_{all}(t)$ at $Z=0.006, 0.008$ and 0.010.
Firstly, the PDFs obtained are all highly non-Gaussian. 
Consequently, it is not appropriate to use the mean of the distribution as the best solution and to quote 1$\sigma$ errors.
Nonetheless, a peak in the PDFs is consistently seen at log(age/years)=6.7 to 6.9 (\about 5 -- 8 Myrs), which contains over 50 per cent (and up to 83 percent) of the distribution in all metallicity and cluster combinations shown in Figure \ref{fig:combined_age_pdf}.

We conclude that the BPASS HRDs favour an age of \about 5 -- 8 Myrs for D118 and D119.
This estimate will be further tested and refined in the following sections. 

Additionally, we note that the combined age of D118 and D119 seems sensitive to the metallicity whereby higher metallicities drive ages down.
We suspect that this is the result of the effects of metallicity on the opacity of stellar atmospheres. 
The increased line-driven opacity at higher metallicity in O stars leads to a more bloated, and therefore cooler and redder envelope. Consequently, younger O stars at high metallicity can look as cool as older O stars at lower metallicities. 

\section{Stellar numbers and ionizing flux}
\label{sec:sec4}
\subsection{Wolf-Rayet to O star ratio}
\label{sec:WR/O}



\begin{table}
\centering
\caption{\label{tab:mass_counts} BPASS predicted total mass, total star count  and  WR/O star ratio for the D118 and D119 clusters at the preferred ages identified in Section \ref{sec:region_ages} and the metallicity of NGC 300 ($Z=0.006$). These predictions were made by comparing the observed number of WR+O stars with luminosity $>10^{4.9} L_{\odot}$ (6 for both D118 and D119) to number of WR+O stars with the same luminosity cut-off in BPASS. For more detail see Section \ref{sec:WR/O} } 
\begin{tabular}{c c c c}
\hline
Age & $\log(M_{\rm tot})$ & Star Count & WR/O  \\
log(years)  & \msol & ($\times10^3$) &  \\
\hline
6.7 & 3.62$^{+0.15}_{-0.23}$ & 3.6$\pm1.5$ & 0.08 \\ 
6.8 & 3.80$^{+0.15}_{-0.23}$& 5.5$\pm2.2$  & 0.15 \\
6.9 & 4.30$^{+0.15}_{-0.23}$ & 17$\pm7$  & 0.73\\
\hline
\end{tabular}
\end{table}

Based on the age of our systems, as determined in Section \ref{sec:region_ages}, we can search within our models to retrieve the number of O stars and WR stars we would expect to find.
By comparing this to the number of O+WR stars observed we can deduce the mass of the system by scaling the standard BPASS synthetic population total mass (10$^6$ \msol).
Additionally, we can independently calculate an age estimate by comparing the observed and expected WR to O star ratios.
Note that here we only consider classical WR stars.

We focus on log(age/years) = 6.7, 6.8 and 6.9, as they were identified as the most likely ages for the D118 and D119 in Section \ref{sec:region_ages}, and summarise the total stellar mass and count obtained for the metallicity of NGC 300 ($Z=0.006$) in Table \ref{tab:mass_counts}.
We also provide the WR/O ratio for these ages. 

It was important to take into account in these calculations that the observed number of stars is likely to be incomplete, as mentioned in \citetalias{mcleod20}.
In order to mitigate the effects of incompleteness we choose to only compare the brightest members of the population.
The BPASS outputs are provided separately for stars with log(L)$\ge$4.9 and stars with log(L)$<$4.9.
Since observations are less likely to suffer from incompleteness at the bright end of the luminosity range, we only consider stars with log(L)$\ge$4.9.


In both D118 and D119 four O stars meet this criteria.
The two WR stars reported with stellar parameters in table 4 of \citetalias{mcleod20} also have a high luminosity and it is safe to assume that the two WR stars in the aforementioned table without reported stellar parameters also belong the the high metallicity category.
This would traditionally result in an observed WR/O = 0.50 $\pm 0.43$ for both D118 and D119, assuming Poisson error. 
This approach to error calculation can be problematic however  (see \citealt{neugent12})  and we instead use a new method presented by \cite{dornwallenstein20} that derives the underlying stellar count ratio from the observed one. 
A summary of this method is provided in the Appendix \ref{app:ratio}.
The underlying WR/O ratio for D118 and D119 is found to be 0.52$^{+0.10}_{-0.22}$.
We can then find the age most likely to produce this WR/O value, we linearly interpolate the BPASS WR/O function between the time bins and find log(age/years) = $6.86^{+0.05}_{-0.06}$.

We can use a similar interpolation technique to infer the cluster mass corresponding to our age estimate, and we find  $\log(M_{\rm tot})=4.18^{+0.39}_{-0.51}$.
The large uncertainties are an inevitable consequence of the errors on the mass (as reported in Table \ref{tab:mass_counts} and on the age estimates.
We note that although stochastic sampling of the IMF can be a significant source of systematic error for low cluster masses (\about 100\msol), above \about 10$^3$ \msol\, the effects of stochasticity are limited as demonstrated by \cite{eldridge12} especially in stellar populations including binary stars.
Consequently, for the masses considered here, the systematics associated with IMF sampling will be negligible compared to the uncertainties resulting from low number statistics.

\subsection{Ionizing flux}
\label{sec:ion}
One of the outputs of the BPASS models is the ionizing flux of the synthetic stellar population. 
This offers an independent method to derive a mass estimate, by comparing the simulated ionizing flux values at our derived age to that observed by \citetalias{mcleod20}. 
The ionizing photon flux ($\log Q_{\rm H\alpha}$) is reported in table 6 of \citetalias{mcleod20} for the individual H\,{\sc ii} regions 118 A, B and 119 A, B, C. 
In order to compare our new mass estimate to the one presented in Section \ref{sec:WR/O}, we combine the values given for the individual H\,{\sc ii} regions to obtain the ionizing flux of D118 and D119 and find $\log Q_{\rm H\alpha}$ = 49.95 s$^{-1}$ and 50.27 s$^{-1}$, respectively. 

To compare these observations to our models we select the output corresponding to the BPASS metallicity $Z=0.006$, which is closest to that of NGC 300. 
Similarly to Section \ref{sec:WR/O}, we interpolate linearly the BPASS ionizing flux between the available time bins in order to recover the value corresponding to log(age/years) = $6.86^{+0.05}_{-0.06}$ Myr.
We then take the ratio of the observed to the simulated ionizing flux and scale the total BPASS stellar population mass (10$^6$\msol) accordingly.

Using this method we find $\log(M/$\msol$)=4.36^{+0.2}_{-0.18}$ and $4.68^{+0.2}_{-0.18}$ for D118 and D119, respectively. 
The large uncertainties on these mass estimates are a direct result of the small number of bright stars observed in the clusters, resulting in large Poisson noise which propagates to the age estimate in Section \ref{sec:WR/O}, and to the values derived here. 

Overall we can conclude that the mass estimates for D118 and D119 derived from the ionizing flux measured by \citetalias{mcleod20} are consistent with the masses we inferred from the number of bright WR and O stars observed (see Table \ref{tab:mass_counts}).
This is a key validation test of the BPASS models. Here we have independently tested for consistency in the code predictions on ionizing flux per number of O stars and the mass of stars formed. 
The excellent agreement between the parameters derived from these two independent measures of the massive star population confirms  that the predictions can be used in other similar studies such as \cite{xiao18, xiao19}

\subsection{BPT diagrams}
A key way insight can be gained into the ionizing radiation spectral energy distribution of a H\,{\sc ii} region derives from the ratios of the mostly-forbidden metal lines that are observed in an optical spectrum of the region in addition to those of the Balmer series of hydrogen. 
These provide information on the parameters of the ionized gas as well as the ionizing spectrum of the stars.
We use the integrated line intensities from the MUSE data of individual H\,{\sc ii}  regions in D118 and D119 from \citetalias{mcleod20} to calculate the lines ratios used in the \citet[][BPT]{1981PASP...93....5B} diagram and compare these to the BPASS predictions of nebular emission for a population of the age we infer above. 

\begin{figure*}
	\includegraphics[width=15cm]{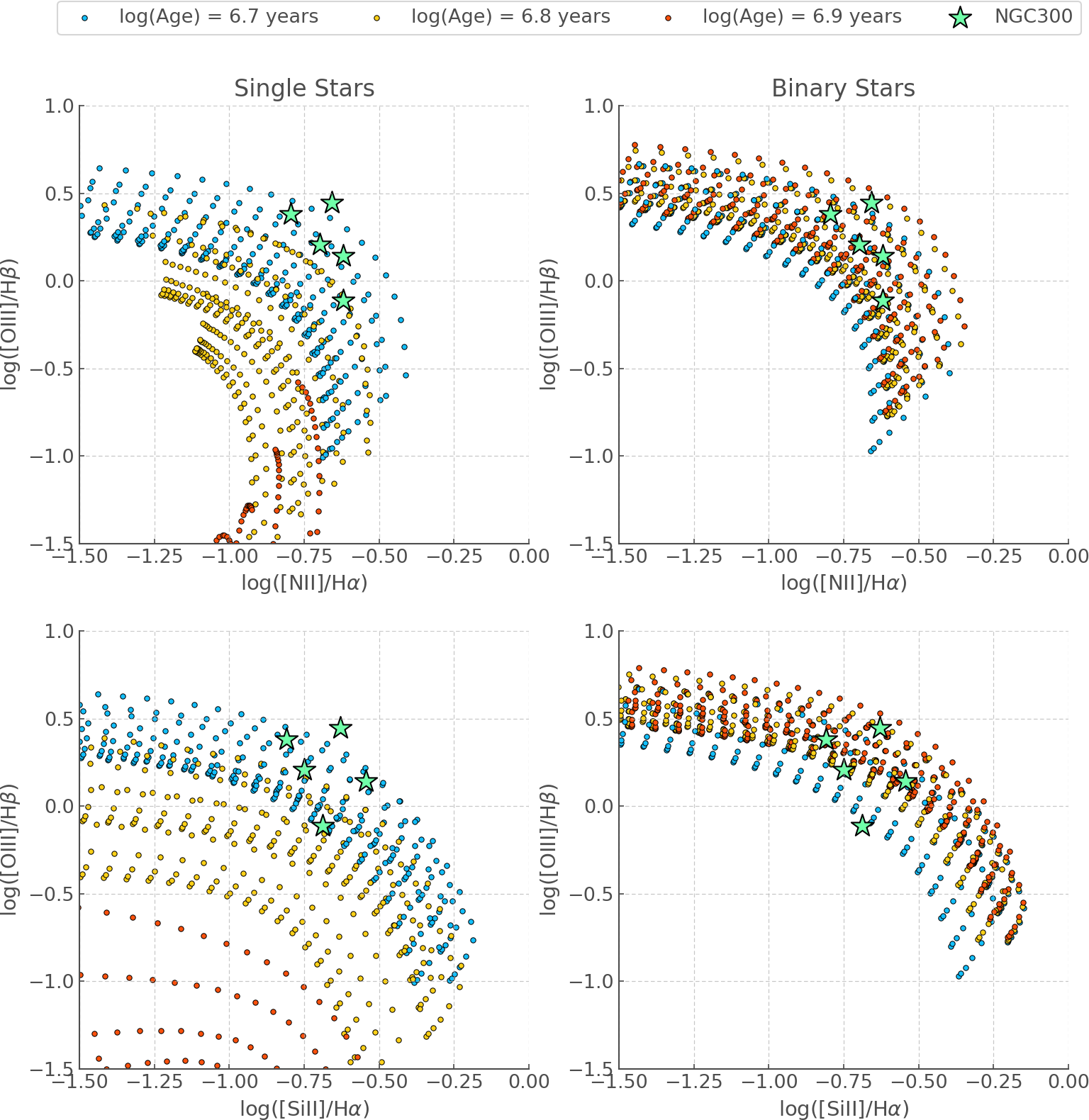}
    \caption{\label{fig:bpt} BPT diagram showing BPASS predicted lines ratios from Xiao et al. 2018 and the line flux ratios for the H\,{\sc ii} regions 118A, 118B, 119A, 119B and 119C. The observed line ratios are shown by green stars. The left-hand panels are for single star population models and the right-hand panel for the binary star predictions. For these tracks the ionization parameter, $U$, increases from left to right and the higher tracks of the same colour represent higher densities for the hydrogen gas around the stars. Note we show models for Z=0.006 and Z=0.008 because they encompass the metallicity of NGC 300.}
\end{figure*}

Figure \ref{fig:bpt} demonstrates that the BPASS models are able to reproduce the observed line ratios at ages inferred for the regions above. However we must note that there is considerable degeneracy in the age for the binary populations. This is because binary interactions lead the ionizing flux from the stars to last for extended periods of time as described by \cite{xiao18, xiao19}. In addition, binary interactions create more hot WR stars at later times,  which results in a less rapid  evolution in the nebular line ratios than would be expected for a single-star population \citep{gotberg19}.
By comparison, for the single star populations, we see that only models with log(age/years)=6.7 are able to match the observed line flux ratios and that this only occurs at the highest hydrogen gas densities.

With a limited number of observed lines, performing a formal fit to the line ratios would lead to significant uncertainty in age and is of limited value. 
By inspection, we find that the ionization parameter $U$ (see Eq. 1 of \citealt{xiao18}) has values in the range of -2.5 to -3 and the gas density of the regions in the range of $log(n /{\rm g \, cm^{-3}})=$2 to 3. Part of the problem in deriving the age of the H\,{\sc ii} regions from the line ratios is that younger stellar populations can also reproduce the line ratios of these regions. However at such ages the stellar populations predict no Wolf-Rayet stars or that the number of O stars significantly outnumber them, which is clearly not the case here. Thus we can discount such interpretations.

Our main finding here however is that our predicted nebular emission line ratios at the age of the stellar population derived from the resolved stellar populations is consistent with that observed. 

This study also supports the interpretation offered by \cite{xiao18} that some H\,{\sc ii} regions observed in other galaxies may in fact be significantly older than has hitherto been assumed, based on the truncation of ionizing flux from a single star population at ages of a few Myrs.

\section{Discussion}
\label{sec:discussion}

\subsection{Preferred Age for D118 and D119}
In Section \ref{sec:region_ages} we deduced ages between 6.7 and 6.9 for D118 and D119, based on the HRD locations of the individual cluster members. 
In Section \ref{sec:WR/O}, however, we estimated log(age/years)=$6.86^{+0.05}_{-0.06}$.

\begin{figure}
	\includegraphics[width=\columnwidth]{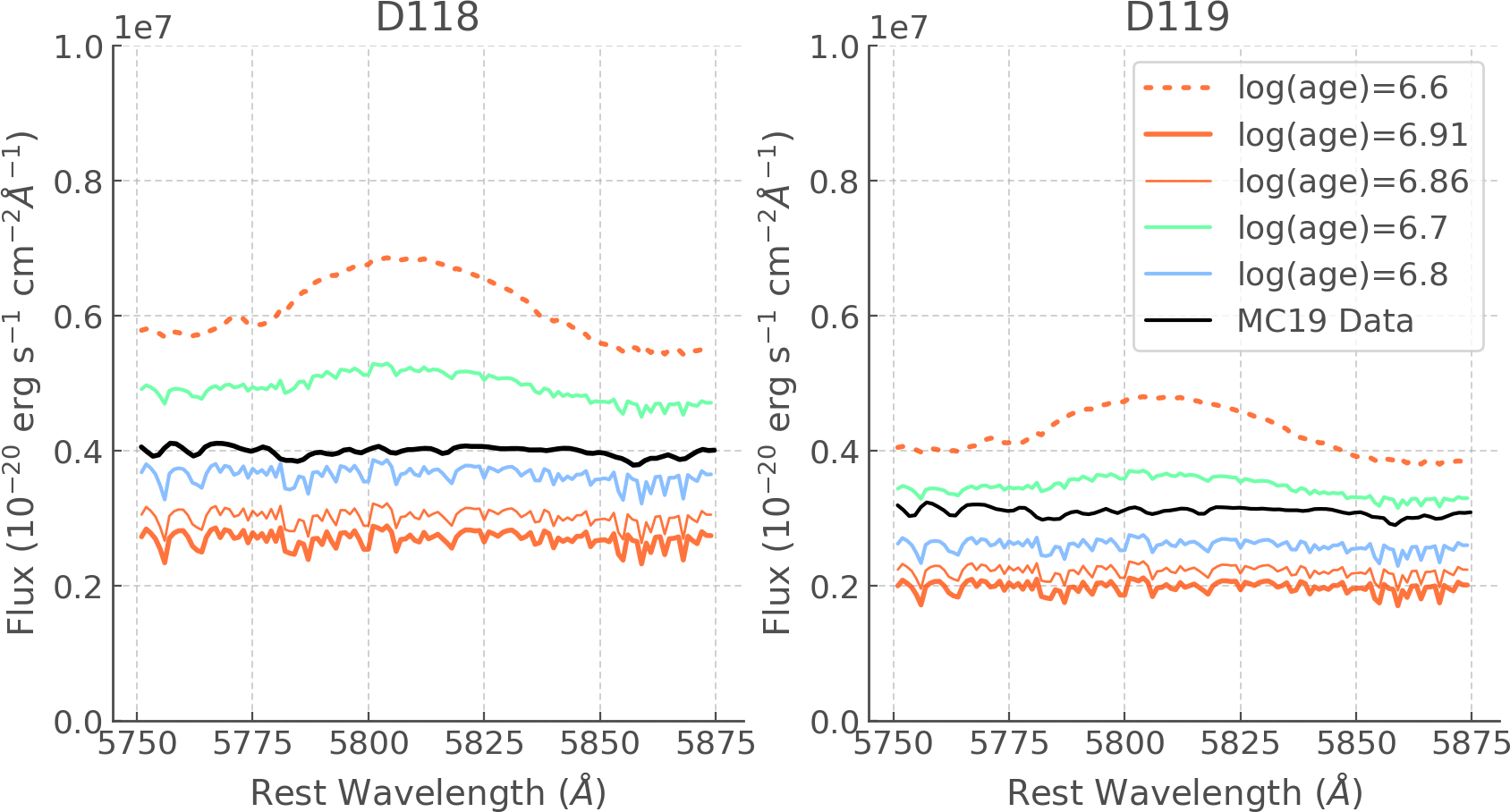}
    \caption{\label{fig:red_bump} Comparison of the spectra of D118 and D119 to BPASS predictions for a range of ages. The BPASS spectra have been scaled to match the units of the \citetalias{mcleod20} observations by visually aligning the red parts of the visible spectrum. The figure was subsequently cropped to the desired range.}
\end{figure}

To further validate the ages of D118 and D119 we can use their spectra as reported by \citetalias{mcleod20} and compare them to BPASS synthetic spectra. 
In particular we focus on the Red Wolf-Rayet bump in the range 5750--5900\ang as the spectral behaviour of the spectrum in this region is highly age dependent \citep{eldridge09}.
In Figure \ref{fig:red_bump} we show the BPASS spectra for our preferred age and its upper and lower boundary (as calculated in Section \ref{sec:WR/O}), as well as all BPASS time bins covered in this time range. 
As we can see, the observed spectrum of D118 and D119 \citepalias{mcleod20} do not show a Red Bump whereas the BPASS spectra of log(age/years) $\le$ 6.7 predict a noticeable feature. 

This corroborates our age estimates calculated in earlier Sections. 
The best derived age (log(age/years)=$6.86^{+0.05}_{-0.06}$) is significantly older than previous estimates by \cite{faesi14} based on comparisons of \halpha and Far UV luminosities to Starbust99 models \citep{claus99}, as they found log(age/year) of 6.5-6.6 for D118 and 6.4-6.5 for D119. 
Although it is known that binary populations (not included in Starburst99 in this earlier study) can make older populations look younger, direct comparison of single star and binary BPASS models for D118 and D119 (see Section \ref{sec:sin_vs_bin}) seems to indicate that in this particular case the use of single star model cannot solely explain such an age discrepancy. 

\subsection{Mass estimate for D118 and D119}

Mass estimates for D118 and D119 were found by comparing their WR/O and ionizing fluxes to those predicted by BPASS (see Section \ref{sec:WR/O} and \ref{sec:ion}, respectively). 
Interpolation between the BPASS time bins was required in order to match the preferred age estimates.
Due to uncertainties in the masses associated with each time bin, and the uncertainties associated with the inferred age, the range of possible masses is quite broad. 

Another way in which we can estimate the masses of D118 and D119 is to instead consider ANGST data \citep{dalcaton09} which includes lower mass stars.
The large number of stars in this data set allows us to construct a magnitude histogram, which we can then compare to BPASS predictions. 
In order to mitigate the effect of an incomplete sample, we only consider the brightest stars, with absolute magnitudes $\le-4$ in F435W.
We can then infer the mass of D118 and D119 by scaling the BPASS prediction to match the observations. 
That scale factor can be applied to the 10$^6$\,\msol\ synthetic population simulated in the models to retrieve our estimate, as we did with the scale factor deduced from the WR/O ratios in Section \ref{sec:WR/O}.

\begin{figure}
	\includegraphics[width=\columnwidth]{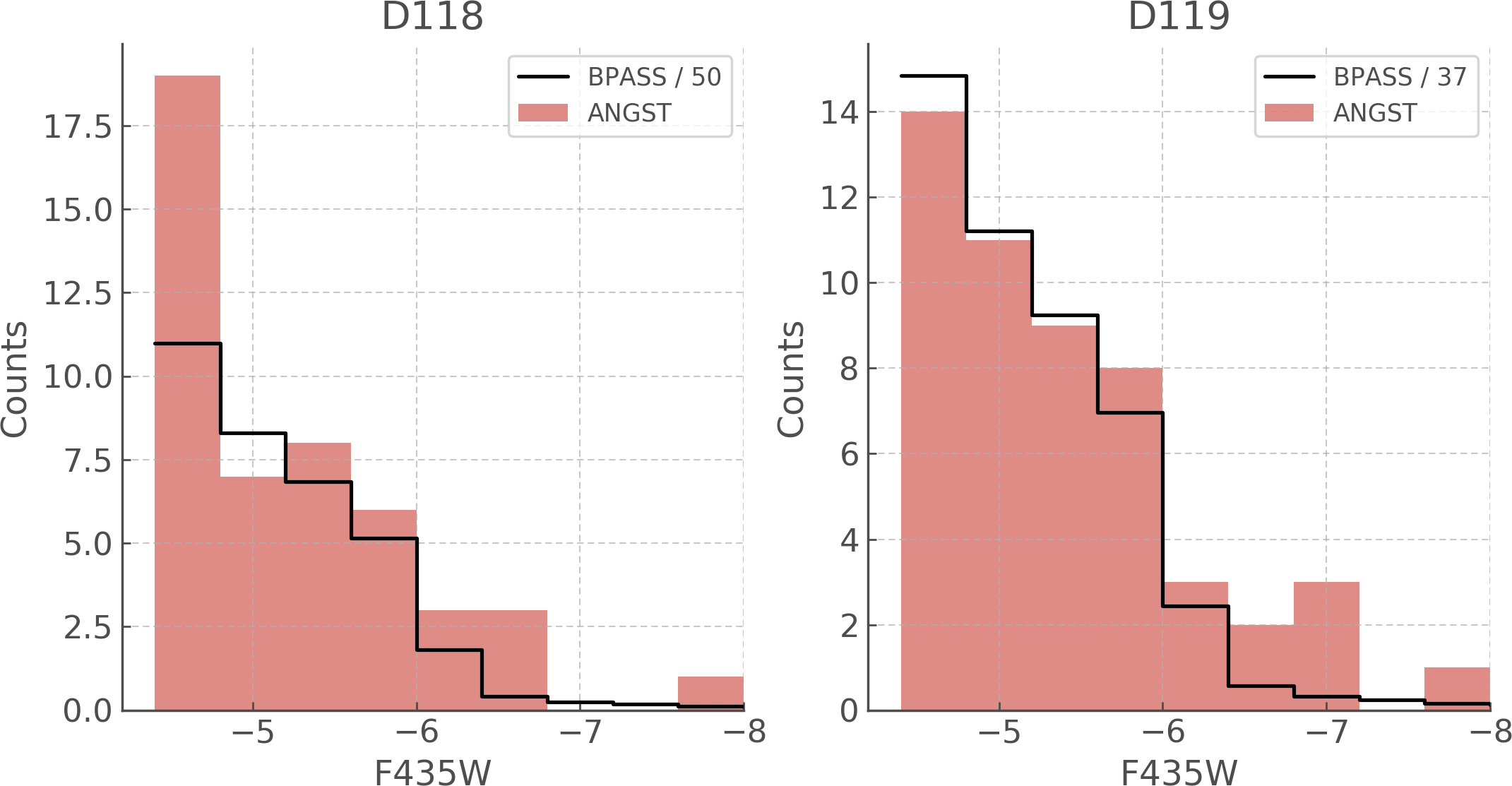}
    \caption{\label{fig:hist}Number of stars with magnitudes $\le -4$ in the ANGST samples of D118 and D119 compared to scaled predictions of BPASS.}
\end{figure}

As a first approach, the scaling was done by visual examination, see Figure \ref{fig:hist}.
The shape of the observed and modelled samples are consistent overall, although D118 shows an excess of fainter stars. 
These could be associated with underlying older populations.
Considering the scaling shown here, we infer masses for D118 and D119 of log(M/\msol)\about4.30 and log(M/\msol)\about4.43, respectively. 
This is consistent with previous estimates in the case of D118, whereas for D119 our mass is 0.2 to 0.4 dex larger than found by \cite{faesi14}.

\subsection{Isochrones versus {\tt hoki}}
As mentioned in the introduction, one of the most common ways of determining the age of an H\,{\sc ii} region is through isochrone fitting on an HRD or CMD.
The age of individual stars can also be deduced if spectroscopy has been obtained. 

Here we compare the well-known Geneva stellar evolution isochrone models to BPASS. 
The Geneva models\footnote{https://www.unige.ch/sciences/astro/evolution/en/research/geneva-grids-stellar-evolution-models/} have been a staple of astronomy for the past 30 years and continue to be refined (e.g. \citealt{schaller92}) and they now include the complex physics of rotation and magnetic field (e.g. \citealt{meynet03, hirschi05}). 
One  important aspect they do not take into account is the effect of binary interactions. 

\begin{figure}
	\includegraphics[width=\columnwidth]{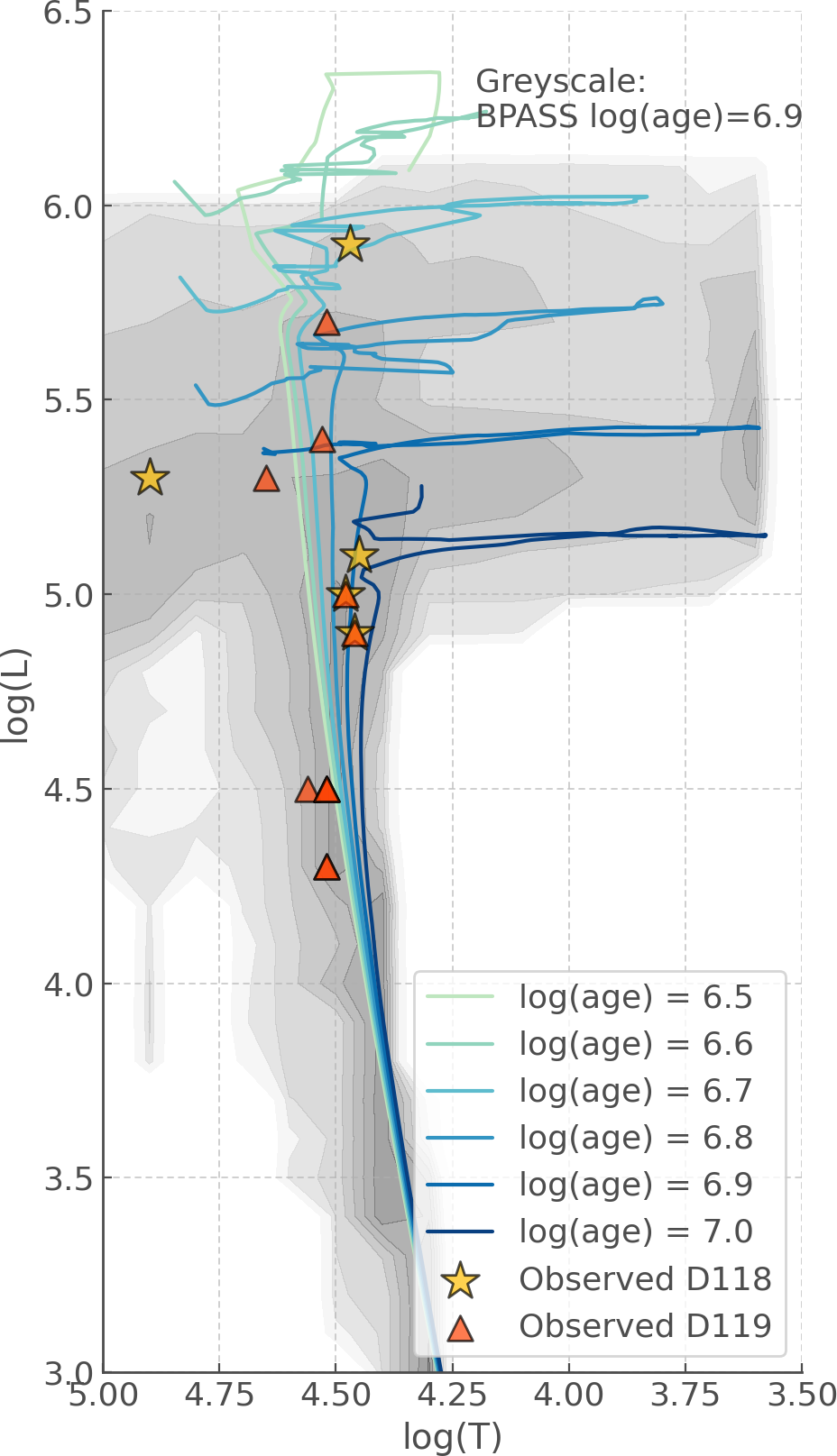}
    \caption{\label{fig:geneva_vs_bpass} HRD showing the BPASS models at log(age/years)=6.9 in greyscale and a set of isochrones from the rotating Geneva models in blue, as well as the D118 and D119 cluster members.}
\end{figure}

In Figure \ref{fig:geneva_vs_bpass} we compare the most recent set of Geneva isochrones to a BPASS HRD at log(age/years)=6.9 (nearest to our preferred log(age/years) of 6.86) and the deduced physical parameters of the D118 and D119 cluster members.

First we note that given the relatively small number of stars, it would be very difficult to provide an age estimate using isochrones alone. 
Additionally, since there is no visible main-sequence turn-off, if we used the position of the brightest cluster members to constrain the age of D118 and D119, we might deduce they are closer to log(age/years)=6.7 or 6.8 and exclude the older ages we find more likely with BPASS. 

Generally speaking, the BPASS models at log(age/years)=6.9 predict a significant number of stars in the upper parts of the HRD which would, within the single star paradigm, be recognised as a sign of younger stellar populations. 
The most luminous stars in BPASS at any age result from either stellar mergers or mass-transfer onto a companion: both scenarios increase the effective initial mass and luminosity of the star.
These blue stragglers could cause observers to underestimate cluster ages by 0.2 dex, which in this age range corresponds to \about 3 Myrs. 
On the scale of the lifetime of a massive star, that is a significant difference. 

\subsection{Single versus Binary}
\label{sec:sin_vs_bin}
\begin{figure*}
	\includegraphics[width=15cm]{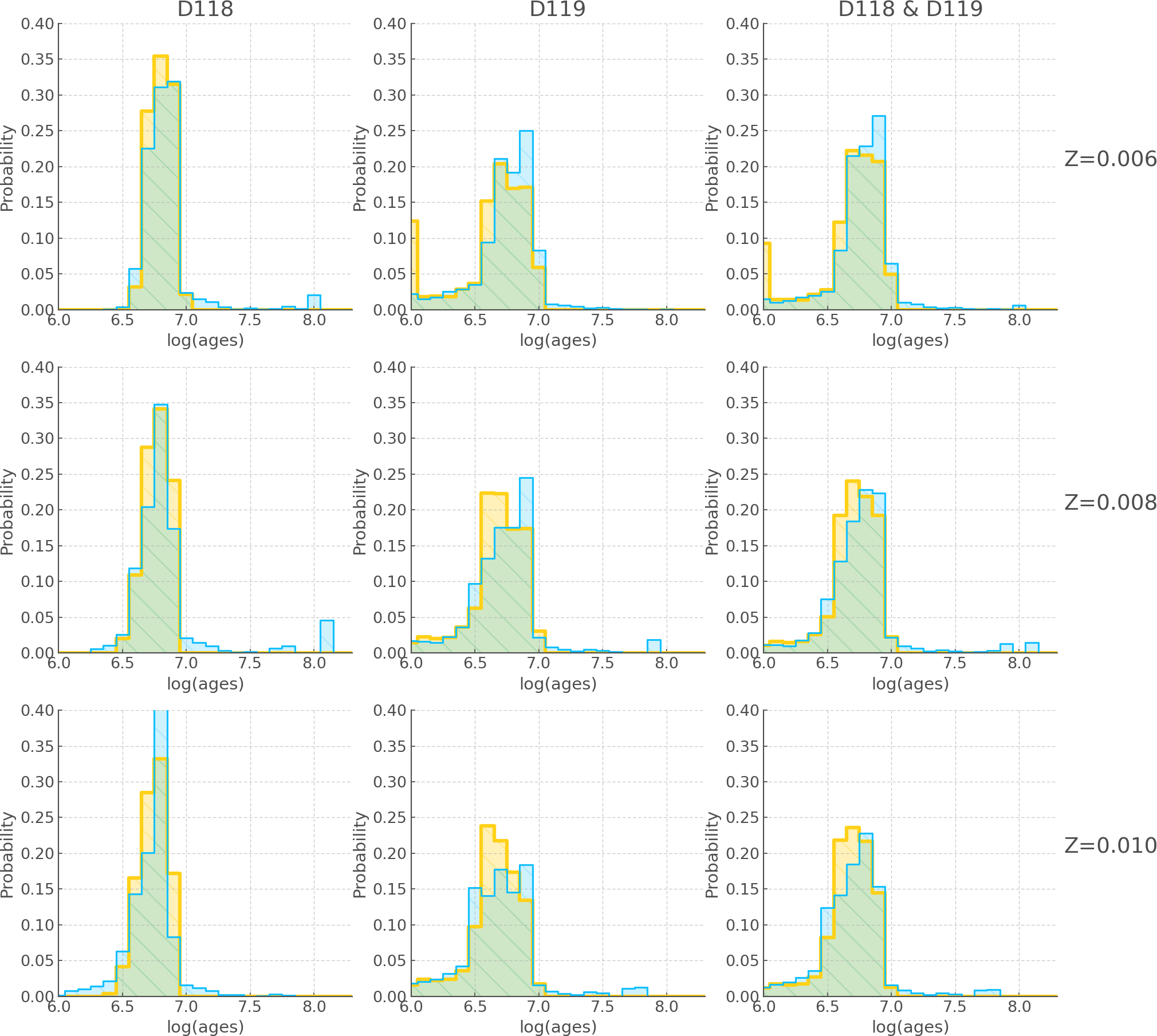}
    \caption{\label{fig:age_sinbin} Comparison of the probability density functions of the ages of D118 and D119 from single star (yellow) and binary star (blue) models. }
\end{figure*}

In order to directly compare the performance of binary and single star models independently of method specific disparities, we create age PDFs for each target as presented in Section \ref{sec:stellar_ages} and present side to side the results from the BPASS population containing solely single stars and from the population that includes binaries (see Figures \ref{fig:agepdfs_sinVsbin}, \ref{fig:agepdfs_sinVsbin2}, \ref{fig:agepdfs_sinVsbin3}).
We summarise the most likely age and the probability that a star has an age between log(age/years)=6.7 to 6.9 in a similar fashion to Table \ref{tab:obs} (see Table \ref{tab:age_sin}), and aggregate the results for our individual sources to find age estimates for D118 and D119 (see Figure \ref{fig:age_sinbin}).

\begin{table*}
\caption{\label{tab:age_sin} Most likely age of individual sources in D118 and D119 and probability the log(age/years) $\in \{6.7,6.8,6.9\}$ inferred from BPASS single star models.}

\begin{tabular}{l c c c r r r}
\hline
ID & \multicolumn{3}{c}{Most likely age} & \multicolumn{3}{c}{P($6.7\le$log(age/years)$\le6.9$)}\\ & \multicolumn{3}{c}{log(years)}\\
 & $Z$=0.006 & $Z$=0.008 & $Z$=0.010 & $Z$=0.006 & $Z$=0.008 & $Z$=0.010  \\
\hline
[DCL88]118-1  & 6.8  & 6.8  & 6.8  & 99 \% & 89 \%  &  79  \% \\ 

[DCL88]118-2  & 6.8 & 6.8 & 6.8 & 100 \% & 94 \%  & 85 \%  \\

[DCL88]118-3  & 6.9 & 6.9 & 6.8 & 86 \% & 78 \%  & 72 \%  \\

[DCL88]118-4  & 6.5 & 6.5 & 6.5 & 1 \% & 0 \%  &0 \%  \\

[DCL88]118-WR2 & -- & -- & -- & -- & --  & -- \\

[DCL88]119-1 & 6.8 & 6.8 & 6.8 & 99 \% & 89 \%  & 79  \% \\

[DCL88]119-2 & 6.7 & 6.7 & 6.7 & 99 \% & 87 \%  & 72 \% \\

[DCL88]119-3 & 6.9 & 6.9 & 6.8 & 46 \% & 49 \%  & 41 \% \\

[DCL88]119-3b & 6.6 & 6.6 & 6.6 & 20 \% & 7 \%  & 4 \% \\

[DCL88]119-4 & 6.9 & 6.9 & 6.9 & 47 \% & 49 \%  & 51 \%  \\

[DCL88]119-5 & -- & -- & -- & -- & -- & -- \\

[DCL88]119-6 & 6.9 & 6.9 & 6.8 & 86 \% & 78 \%  &  72 \% \\

[DCL88]119-7 & 6.9 & 6.9 & 6.9 & 47 \% &  49 \%  & 51 \% \\

[DCL88]119-8 & 6.9 & 6.9 & 6.8 & 46 \% & 49 \%  & 41 \% \\

[DCL88]119-9 & 6.9 & 6.9 & 6.9 & 47 \% & 49 \%  &  51 \% \\

[DCL88]119-WR1 & 6.0 & -- & -- & 0 \% & -- &  -- \\
\hline

\end{tabular}
\end{table*}

This direct comparison of binary and single star models shows three primary results which we discuss in the following subsections. 

\subsubsection{The missing stars}
The single star models are not able to find solutions for all stars in our sample, contrary to the binary models. 
118-WR2 and 119-5 have no solutions at any of the metallicities considered here.
119-WR1 only has a solution at Z=0.006 but given that the resulting distribution is only populated in the log(age/years)=6.0 bin with a 100 per cent probability and a 0 per cent probability at all other ages, we do not consider this to be an accurate representation of reality. 

On the whole, the single star models are not able to reproduce 20 per cent  ($\pm$10 per cent\footnote{Note the error given here is the Standard deviation of the Binomial Distribution. For a sample size of 15 and a probability of 0.2, the binomial distribution is effectively symmetrical within the quoted precision.}) of our sample. 
Even for larger samples we can expect single star models to be unable to predict a significant fraction of the population.

\subsubsection{Consistent preferred ages}
For the stars which have a solution in the single star models the preferred age is the same or very close (within 0.1 dex) of that found in the binary models. 
This is to be expected for stars that would not have interacted yet as they would be found on the HRD at the same location as stars that have no companion.
For the case of 118-4, the single star model also finds a lower preferred log(age/years)=6.5, consistent with our interpretation that the rejuvination models in BPASS should be revisited.

\subsubsection{Narrower age distributions}
As seen in Table \ref{tab:age_sin}, the probability of a source having log(age/years) $\in \{6.7,6.8,6.9\}$ is higher in single star models than in binary models in \about 2/3 cases. 

This indicates that the age distributions for single star models are narrower than in binary models, which is consistent with our current understanding of the effects of binary interactions on stellar populations.

\section{Conclusions}
\label{sec:conclusion}

In this work we presented an ageing method based on the comparison of BPASS models to observation using \hoki.
We used the physical parameters derived by \citetalias{mcleod20} for sources in the groups of H\,{\sc ii} regions D118 and D119 located in NGC 300 to estimate their ages, cluster mass, and stellar count.
This data set also allowed us to verify the consistency of the BPASS models through comparing different predicted observables which lead to the same conclusions. 

Our main findings are as follows: 
\begin{enumerate}
    \item The method introduced in Section \ref{sec:mymethod} allows us to find age estimates for young clusters even with a low sample size. 
    \item For D118 and D119, we find log(age/years)=$6.86^{+0.05}_{-0.06}$  
    \item We highlighted some of the difficulties of using binary models: Very young looking stars like 118-4 could be merger products, whereas stars with age PDFs peaking at later ages may nonetheless be younger main sequence stars in the tail of the distribution. This indicates that BPASS may need to consider its rejuvenation and that extreme age outliers can be understood through manual checks of the underlying models. 

    \item We simultaneously validated multiple observable quantities predicted by the BPASS models: the number counts, stellar type ratios, stellar mass, ionizing flux and BPT diagrams.
        
    \item A comparison of our method to the isochrone fitting technique revealed that the ages of H\,{\sc ii} regions like those in NGC 300 could be underestimated by 0.2 dex (in this context \about 3 Myrs). 
        
    \item A direct comparison of the BPASS single star models with their binary counterpart showed that 20 per cent of our sample had no valid solution. The preferred ages for the sources that did have a solution were similar (within 0.1 dex) of those found with binary models. The age PDFs of single star models were also somewhat narrower than for bianry models.
\end{enumerate}

Based on this study of NGC 300 we propose that potential inaccuracies with isochrone fitting would not solely be the result of their lack of binary star modelling and that there are fundamental limitations to the method itself. 
Nevertheless, the inability of single star models to predict a significant fraction of our population (20 percent $\pm 10$ per cent here) suggests that binary interactions should be taken into account when possible. 

To allow the community to implement this ageing method effortlessly, we have released \hoki v1.5 ("The {\tt AgeWizard} release") which includes all the necessary tools to infer likely ages from comparison of observational data with BPASS HRDs and CMDs.

In the next papers of this series we plan to further refine and test this method by applying it to larger samples, older clusters, and showing the CMD applications of this technique. 

\section*{Acknowledgements}
The authors would like to thank the referee for his careful review of our work. 
HFS and JJE acknowledge the support of the Marsden Fund Council managed through Royal Society Te Aparangi. ERS receives support from United Kingdom Science and Technology Facilities Council (STFC) grant number ST/P000495/1

\textbf{Software:} matplotlib \citep{matplotlib}--  Numpy \citep{numpy} -- pandas \citep{pandas1, pandas2} -- emcee \citep{emcee} -- corner \citep{corner} --  astropy \citep{astropy13, astropy18}

\section*{Data Availability}

The full data analysis code and numerical data used for this analysis can be found in the Jupyter notebooks we are making available with this publication. You can download them at this link: https://github.com/UoA-Stars-And-Supernovae/A\_systematic\_ageing\_method\_I.
If you struggle running these notebook, feel free to open an issue on the GitHub page or send en email to hfstevance@gmail.com





\bibliographystyle{mnras}
\bibliography{./bib} 

\appendix
\section{Supplementary plots}
\label{appendix}
\begin{figure*}
	\includegraphics[width=14cm]{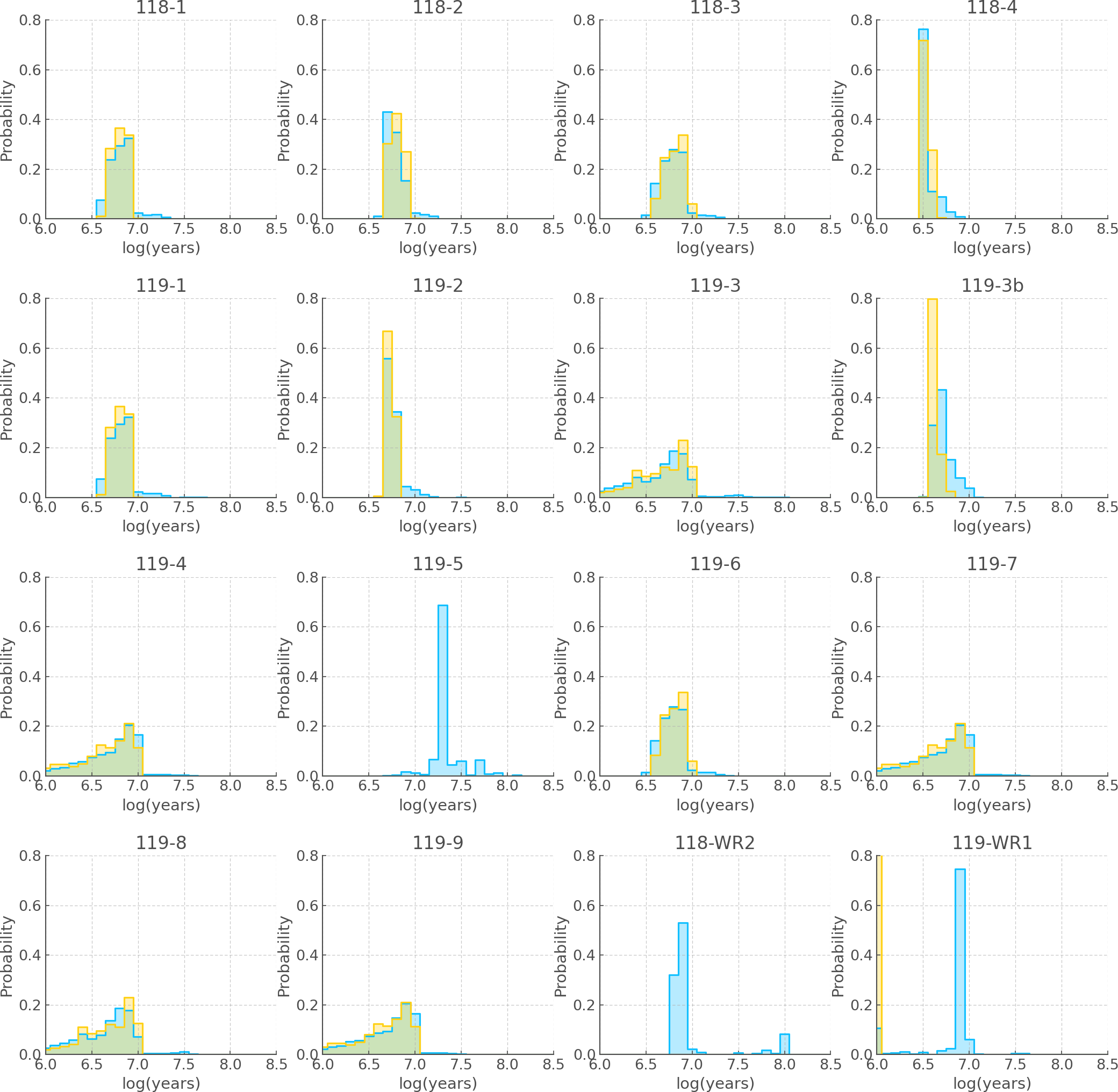}
    \caption{\label{fig:agepdfs_sinVsbin}probability density functions of the age of each star inferred from the stellar parameters quoted in Table \ref{tab:obs} and the synthetic HRDs of BPASS for a standard IMF and a metallicity $Z=0.006$ for single star (yellow) and binary (blue) models.}
\end{figure*}

\begin{figure*}
	\includegraphics[width=14cm]{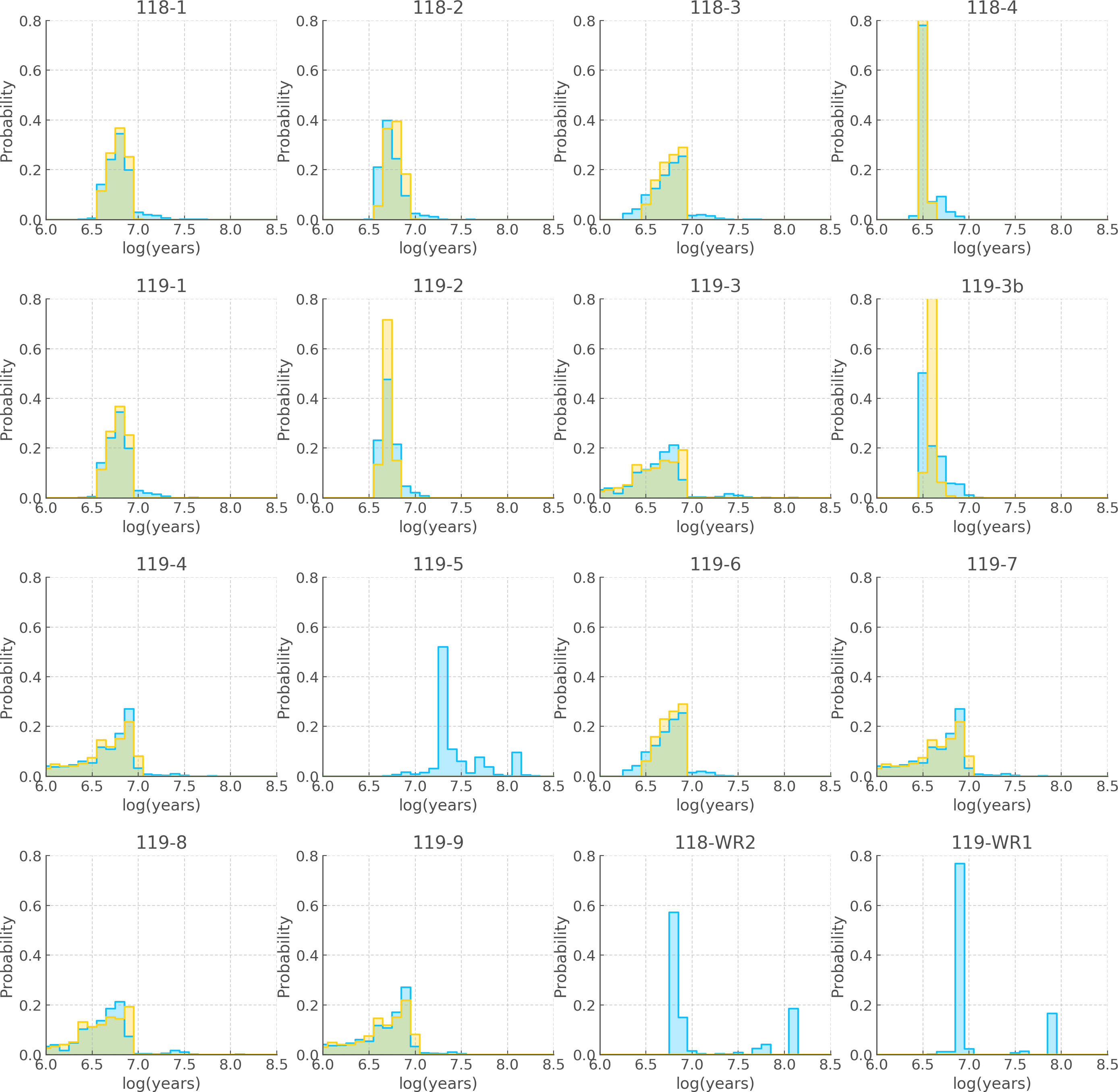}
    \caption{\label{fig:agepdfs_sinVsbin2}Same as Figure \ref{fig:agepdfs_sinVsbin} for $Z=0.008$.}
\end{figure*}

\begin{figure*}
	\includegraphics[width=14cm]{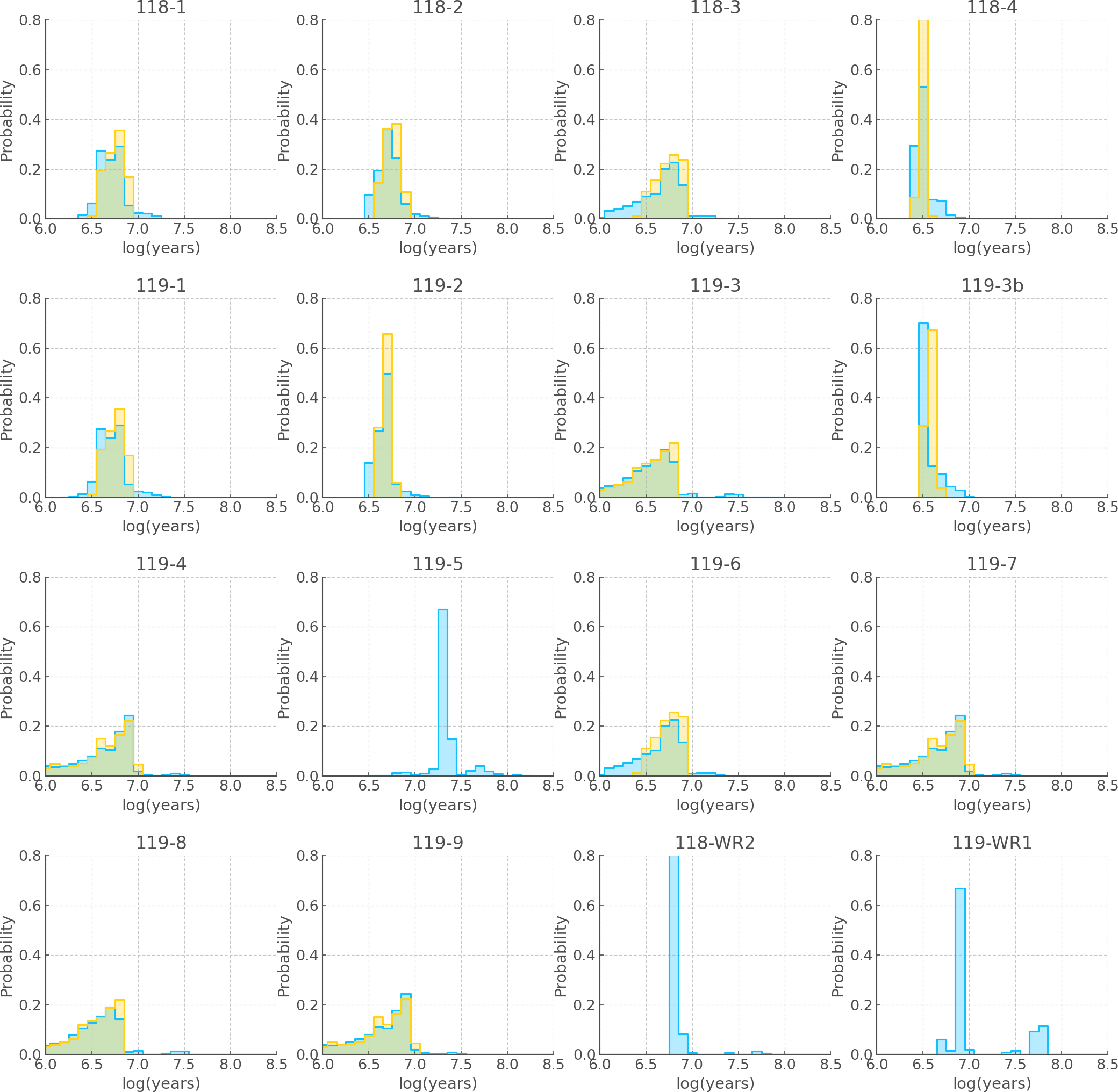}
    \caption{\label{fig:agepdfs_sinVsbin3}Same as Figure \ref{fig:agepdfs_sinVsbin} for $Z=0.010$.}
\end{figure*}

For completeness we show in Figures \ref{fig:agepdfs_sinVsbin}, \ref{fig:agepdfs_sinVsbin2} and \ref{fig:agepdfs_sinVsbin3} a comparison of the age PDFs obtained for binary star models (previously presented in Figures \ref{fig:stars_age_pdf}, \ref{fig:stars_age_pdf_z008} and\ref{fig:stars_age_pdf_z010}) and single star models.

\newpage
\section{WR/O star ratio}
\label{app:ratio}
\subsection{Introduction}
In Section \ref{sec:WR/O} we used the \cite{dornwallenstein20} method to estimate the \textit{underlying} WR/O number count ratio ($\hat{R}$ ) from the observed one ($R$). \cite{neugent12} discussed that although shot-noise (Poisson errors) is a reasonable assumption for stellar number counts, we can expect a low number count ratio to have asymmetrical errors. 
In addition, with few enough stars it is well within a traditional "3$\sigma$" error bar to retrieve an observed ratio $R=\inf$ or 0, which would no reflect the underlying population ratio $\hat{R}$.
\cite{dornwallenstein20} proposed a Bayesian method where, given $\hat{R}=n_1/n_2$, $\hat{R}$ and $n_2$ can be estimated using a Markov-Chain Monte Carlo run. For more detail, see Section 3.2 of \cite{dornwallenstein20} and reference therein.

With permission from Dorn-Wallenstein (priv. correspondence), we have added a tool within \hoki to implement this method. Further details are included in the Jupyter Notebook\footnote{https://github.com/UoA-Stars-And-Supernovae/A\_systematic\_aging\_method\_I}.

\subsection{Results}
\begin{figure}
	\includegraphics[width=10cm]{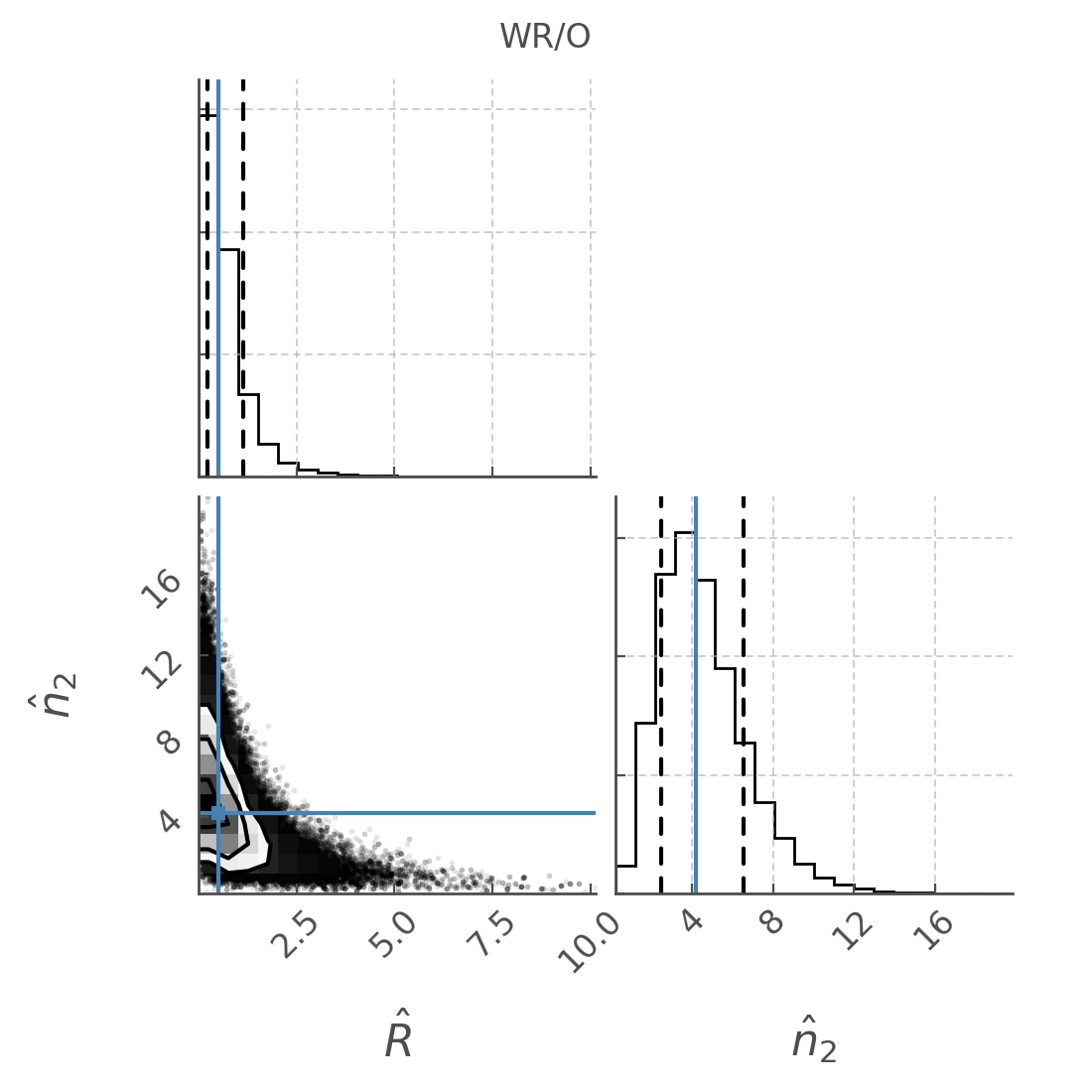}
    \caption{\label{fig:corner} Corner plot showing the marginalised posterior distributions of $\hat{R}$ and $\hat{n_2}$.}
\end{figure}

For both D118 and D119 we record 2 WR stars and 4 O stars above our luminosity cut-off of $log(L/L_{\odot})=4.9$. 
We ran a Markov-Chain experiment with 100 walkers, 500 initial burn-in steps and a 3000 posterior sampling steps. 
The posterior distributions the underlying number count ratio $\hat{R}$ (WR/O in this case) and $n_2$ (i.e. number of O stars) is presented in Figure \ref{fig:corner}.
We also summarise our the 16th, 50th and 84th percentiles in Table \ref{tab:R_n2}.
These percentiles were chosen to mimic the tradition 1$\sigma$ errorbar most commonly used in astronomy. 
The final underlying WR/O star ratio is $0.52^{+0.10}_{−0.22}$ as quoted in Section \ref{sec:WR/O}.

\begin{table}
\caption{\label{tab:R_n2} Summary of the 16th, 50th and 84th percentile for $\hat{R}$ and $\hat{n_2}$ as calculated from the Markov-Chain Monte Carlo simulations. Also see Figure \ref{fig:corner}}
\begin{tabular}{c c c c}
\hline
Variable & 16th &	50th &	84th   \\
\hline
$\hat{R}$ & 0.2983	& 0.5168 & 0.6185 \\
$\hat{n_2}$ &	1.6906 &	2.3822	& 4.1815 \\

\hline
\end{tabular}
\end{table}



\bsp	
\label{lastpage}
\end{document}